\documentclass[usenatbib,preprint]{mn2e}

\usepackage{graphicx}
\usepackage{rotating}

\newcommand{\msol}{\hbox{$M_\odot$}} 
\newcommand{\lsol}{\hbox{$L_\odot$}}         
\newcommand{\e}[1]{$10^{#1}$}
\newcommand{\ee}[1]{$\times 10^{#1}$}       
\newcommand{\kms}{\,km\,s$^{-1}$} 
\newcommand{\cm}[1]{\,cm$^{#1}$}
\newcommand{\ergs}{\,erg\,s$^{-1}$\,cm$^{-2}$\,sr$^{-1}$}
\newcommand{\erg}{\,erg\,s$^{-1}$\,cm$^{-2}$}
\newcommand{\co}{$^{12}$CO}    
\newcommand{\tco}{$^{13}$CO}
\newcommand{\htco}{H$_2$CO}
\newcommand{\hcop}{HCO$^+$}
\newcommand{\oh}{OH(1720 MHz)\ }
\newcommand{\h}{H$_2$}
\newcommand{\mb}{main--beam brightness temperature\ }
\newcommand{\tmb}{$T_{mb}$}
\newcommand{\tkin}{$T_{kin}$}
\newcommand{\tex}{$T_{ex}$}

\newcommand{\nh}{$n_{\scriptscriptstyle\rm H_2}$}
\newcommand{\hone}{2.12\,$\micron$ H$_2$ 1--0 S(1)\ }
\newcommand{\htwo}{2.25\,$\micron$ H$_2$ 2--1 S(1)\ }

\newcommand{\aap}{A\&A}
\newcommand{\aj}{AJ}
\newcommand{\apj}{ApJ}
\newcommand{\apjl}{ApJL}
\newcommand{\apjs}{ApJS}
\newcommand{\mnras}{MNRAS}
\newcommand{\pasj}{PASJ}
\newcommand{\pasp}{PASP}

\begin{document}

\title[Shocked molecular gas towards the SNR G359.1--0.5 and the Snake]
{Shocked molecular gas towards the SNR G359.1--0.5 and the Snake}
\author[Lazendic et~al.]
       {J. S. Lazendic,$^{1,2}$\thanks{Current address: Harvard--Smithsonian Center for Astrophysics, 60 Garden Street, Cambridge, MA 02138, USA} M. Wardle, $^1$ M. G. Burton,$^3$ F. Yusef-Zadeh,$^4$ J. B. Whiteoak,$^2$ \\ 
\newauthor  A. J. Green,$^1$ M. C. B. Ashley$^3$    \\
 $^1$ School of Physics A28, University of Sydney, Sydney NSW 2006, Australia\\
 $^2$ Australia Telescope National Facility, CSIRO, PO Box 76, Epping NSW 1710, Australia \\
 $^3$ School of Physics, University of New South Wales, Sydney NSW 2052, Australia\\
 $^4$ Department of Physics and Astronomy, Northwestern University, Dearborn Observatory, 2131 North Sheridan Road, Evanston, \\ IL 60201-2900 }

\date{Accepted}


\maketitle

\begin{abstract}
We have found a bar of shocked molecular hydrogen (\h) towards the
\oh maser located at the projected intersection of supernova remnant (SNR) 
G359.1--0.5 and the nonthermal radio filament, known as the Snake. The \h\ bar is well aligned with the SNR shell and almost perpendicular to the Snake.  The \oh maser is located inside the sharp western edge of the \h\ emission, which is consistent with the scenario in which the SNR drives a shock into a molecular cloud at that location. The spectral--line profiles of \co, \hcop\ and CS towards the maser show broad--line absorption, which is absent in the \tco\ spectra and most probably originates from the pre--shock gas. A density gradient is present across the region and is consistent with the passage of the SNR shock while the \h\ filament is located at the boundary between the pre--shocked and post-shock regions.
\end{abstract}

\begin{keywords}
supernova remnants  -- ISM: clouds, masers, shock waves -- individual: G359.1--0.5, Snake -- infrared: ISM
\end{keywords}

\section{Introduction}

\oh maser emission detected towards supernova remnants (SNRs) has been
attributed to shock waves driven into adjacent molecular clouds
\citep{frail94}.  The production of this maser emission, without the
simultaneous production of maser emission in the other three ground
state OH transitions  at 1612, 1665 and 1667~MHz, requires specific
conditions (gas density \nh$\sim$\e{5}\cm{-3}, gas temperature
\tkin$\sim$50 -- 125~K, OH column  density $N_{\rm OH}\sim$ \e{15} --
\e{16}\cm{-2}) which can be attained in the cooling gas  behind a
non-dissociative shock wave \citep*{lock99,wardle99}.  Surveys by
\citet{frail96}, \citet{green97} and \citet{kor98} have found that
about 10 per cent  of  $\sim$180 observed Galactic SNRs are associated
with \oh masers. The detection fraction increases to 30 per cent for objects in the Galactic Centre region  \citep{yz96-1}.

Perhaps the most striking result of the survey towards the Galactic Centre
 region is the detection of OH masers along the radio--continuum shell
 of the SNR G359.1--0.5 \citep{yz95,yz96-2}. The brightest of the
 masers (maser {\em A} from \citealt{yz95}) occurs where a nonthermal
 filament known as `the Snake' appears to cross the western edge of
 the SNR shell.  The filament differs from other Galactic Centre
 nonthermal filaments in showing a number of kinks along its 20 arcmin
 extent \citep{grey95}, and its origin is not well established
 \citep{nicholls95,benford97,uchida96}. The latest model by \citet{bi00} proposes that the Snake is a magnetic flux tube anchored in dense rotating material. It has been suggested that the Snake is interacting with the shell of G359.1--0.5 because both objects show change in brightness at the apparent crossing point  \citep{uchida92-2,grey95}. This could, however, be an artifact due to superposition of disparate components along the line of sight. Nevertheless, the presence of an \oh maser -- a signature of SNR/molecular cloud interaction -- at this location is quite intriguing.  In order to investigate the proposed interaction between the Snake and the SNR, and to characterise the ambient molecular gas in the region, we observed molecular hydrogen (\h) and a number of other molecular species, particularly searching for signatures of shocked gas. The observations are discussed in section 2, the results are presented in section 3 and discussed in section 4. The conclusions are presented in section 5.

\section{Observations}

\subsection{UNSWIRF observations}

The observations towards maser {\em A} ($\rm{RA}(1950)=17^{\rm{h}}~41^{\rm{m}}~46 \fs042$, $\rm{Dec.}~(1950)=-29\degr~49\arcmin~51 \farcs03$) were carried out in June 1998 with UNSWIRF\footnote{University of New South Wales Infrared Fabry--Perot \citep{ryder98} } on the Anglo--Australian Telescope (AAT). Two data sets comprising five and six Fabry--Perot frames, each spaced by 40\kms, were taken in the \hone transition. An additional frame was taken at a setting of $-$400\kms\ to enable subsequent continuum subtraction. The integration time was 120 seconds per frame. A resulting image has diameter of 100 arcsec and a pixel size of 0.77 arcsec. The velocity resolution was $\sim$ 75\kms\  FWHM. All the data were reduced using modified routines in the {\sc iraf} software package as described in \citet{ryder98}.  Intensity calibration was performed using the standard star BS~5699.  A velocity cube was constructed by averaging the frames at the same Fabry--Perot setting from two data sets, which was then fitted with the instrumental Lorentzian profile to determine the line flux (to within 30 per cent) and central velocity (to within 20\kms)
for the H$_2$ emission across the field. We note that these data do
not provide a true velocity--spatial cube of the emission, as  the instrument does not have the velocity resolution to provide an accurate line velocity at each position. Rather, the line velocity is the central velocity at each position. Changes in line center velocity can, however, be determined. To confirm the detection, the observations were repeated in June 1999 for the 1--0 S(1) line and additional observations were obtained in the \htwo line. 
On this occasion five Fabry--Perot frames were used for both transitions, with an integration time of 180 seconds per frame. To establish the coordinate scale for the UNSWIRF images, we used the positions of stellar sources in the K-band continuum image listed in the Two Micron All Sky Survey (2MASS) point source catalogue \citep{cutri97} (which has an accuracy of $\approx 0 \farcs1$).

\subsection{SEST observations}

A millimetre--line survey towards maser {\em A}  was carried out  with the 15--m SEST\footnote{The Swedish--ESO Submillimetre Telescope (SEST) is operated by the Swedish National Facility for Radio Astronomy, Onsala Space Observatory and by the European Southern Observatory (ESO).} at La Silla, Chile during June 2000. Position--switching was used with a reference position at $\rm{RA}(1950)=17^{\rm{h}}~45^{\rm{m}}~00^s$, $\rm{Dec.}~(1950)=-31\degr~00\arcmin~00\arcsec$. For periodic pointing calibration and focusing the 3-mm  SiO maser of AH Sco was observed. The main--beam brightness temperature (\tmb) scale of the spectra was calibrated on--line using a black--body calibration source at ambient temperature, and later corrected for the telescope main--beam efficiency (0.74, 0.70, 0.67 and 0.45 at 85 -- 100~GHz, 100 -- 115~GHz, 130 -- 150~GHz and 220 -- 265~GHz respectively). Simultaneous observation of transitions in two different bands was available using dual SIS receivers at 3 and 2~mm, or 3 and 1.3~mm.

  For spectral--line observing we used an acousto--optical spectrometer split into two 43--MHz bands, each with 1000 channels, providing velocity coverages ranging from 60 to 100\kms\ and velocity resolution ranging from 0.06 to 0.14\kms\  across the total wavelength range. A low--resolution spectrometer with resolution of  $\sim$1--2\kms\ was also used simultaneously to determine the spectral baselines in the case of broad--line emission. The final data were smoothed  over three channels. The observed transitions, frequencies and antenna beam sizes (FWHM) are listed in Table~\ref{tab-sest}. For the \co, \tco\ and CS transitions we obtained maps with approximately half--beam and one--beam sampling intervals at lower and higher transitions respectively. The final images covered the $2\times2$ arcmin$^2$ region around maser {\em A} that coincides with the UNSWIRF field of view. The integration time was 30 seconds for CO isotopomers and 90 seconds for CS transitions. A set of spectra on--source and at offsets of 20 arcsec in {\rm RA} and {\rm Dec.} was obtained for each of the other molecular transitions listed in  Table~\ref{tab-sest}, with integration times of 180 seconds at each position.
\begin{table}
\caption{The first column lists observed molecular species, the next three columns list the SEST observational parameters and the last column gives observed \mb (\tmb) and rms noise towards the maser {\em A}.}
\begin{tabular}{@{}lcrcc}
\hline\hline
Molecule & Transition & Frequency & Beam Size & \tmb \\
 &  & (GHz) & (arcsec)  & (K) \\   
\hline
\tco  & 2--1 & 220.399 & 23 & 0.78$\pm$0.21 \\ 
      & 1--0 & 110.201 & 45 & 1.20$\pm$0.26  \\
      &      &           &  &   \\
\co   & 2--1 & 230.538 & 23 & 16.00$\pm$0.20  \\ 
      & 1--0 & 115.271 & 45 & 10.30$\pm$0.50  \\
CS    & 3--2 & 146.969 & 34 & 0.42$\pm$0.12  \\
      & 2--1 & 97.981  & 52 & 0.54$\pm$0.14 \\
\hcop & 1--0 & 89.188  & 54  & 0.48$\pm$0.08  \\
\\
C$^{18}$O & 2--1 & 219.560 & 23 & $<$ 0.16  \\
      & 1--0 & 109.782 & 45 & $<$ 0.12 \\
HCN   & 1--0 & 88.632  & 55  & $<$0.17   \\
\htco & 3$_{(2,2)}$--2$_{(2,1)}$& 218.475 & 24 & $<$ 0.13 \\
      & 3$_{(0,3)}$--2$_{(0,2)}$& 218.222 & 24 &  $<$ 0.15   \\
SiO   & 5--4 & 217.106 & 24 & $<$ 0.14  \\
      & 2--1 & 86.848 & 57 & $<$0.17  \\
\hline
\end{tabular}

\medskip
The first part of the table lists results from Gaussian fit of the two \tco\ transitions, whose line profiles peak at $-13$\kms\ and have line widths of $\sim$15 -- 20\kms. The second part of the table lists the parameters of \co, CS and \hcop\ transitions. Their line profiles suffer broad--line absorption and Gaussian fitting was not possible. We therefore list only the maximum \tmb\ of their line profiles, which occur at $-5$\kms. The third part of the table lists the non--detections. Upper limits
 are 2$\sigma$ values.
\label{tab-sest}
\end{table}

\section{Results and analysis}

\subsection{H$_2$ emission}

\begin{figure}
\includegraphics[height=7cm]{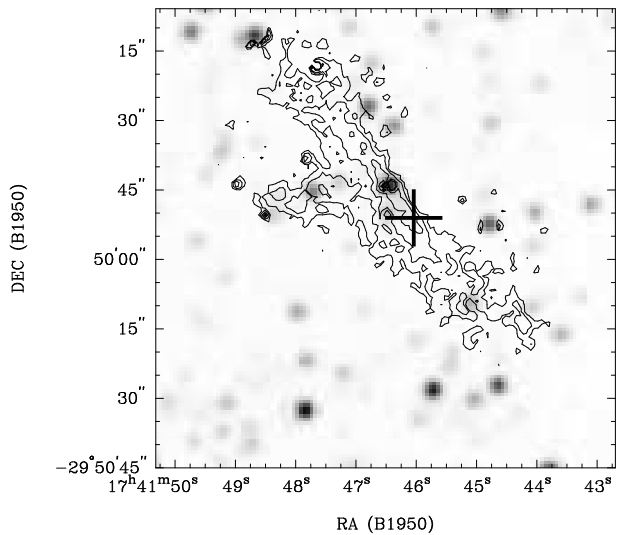}
\caption{Image of the velocity--integrated \hone line (contours) overlaid onto the 2MASS K-band image (greyscale). The contours are: 13, 20, 30, 45, 56 and 65 \ee{-6}\ergs. The cross  marks the location of \oh maser {\em A}. }
\label{fig-h2}
\end{figure}

Fig.~\ref{fig-h2} shows contours of velocity--integrated \hone line 
emission superimposed on a greyscale image of the 2MASS K-band. Some
small--scale structure apparent in the \h\ distribution are residuals
from subtracting the stellar continuum. A bar of H$_2$ emission, 1.5
arcmin in length and 15 arcsec in width, extends from the north--east
to the south--west.  We note, however, that there is weak extensive
diffuse emission across the observed region, which was not included in
fitting procedure because low S/N. The lowest contour in
Fig.~\ref{fig-h2} thus corresponds to a (4 $\sigma$) flux level of
2\ee{-5}\ergs. The peak flux density is 10.6\ee{-4}\ergs\ and the
total flux density is 9.2\ee{-13}\erg.  After correcting for the
typical extinction of the Galactic Centre region (3 magnitudes in
K--band) the parameters of the 1--0 emission are: peak flux density of
1.7\ee{-2}\ergs, total flux density of (1.4$\pm$0.4)\ee{-11}\erg\ and
luminosity of $\sim$33~\lsol. There is no obvious emission in the
individual frames of the \htwo data. However, adding the pixels in the
central frame of the 2--1 cube over the region delineated by the 1--0
emission, we derive a total flux density of (2.4$\pm$0.4)\ee{-14}\erg, which is $\sim$ 2.9\ee{-13}\erg\ after correcting
for extinction. The corresponding total flux density in the 1--0 peak
frame is 3.1\ee{-13}\erg, i.e. 4.9\ee{-12}\erg\ after correcting
for extinction. The ratio between the 1--0 and 2--1 line
emission is thus $\sim 20$.

\begin{figure}
\includegraphics[height=7cm]{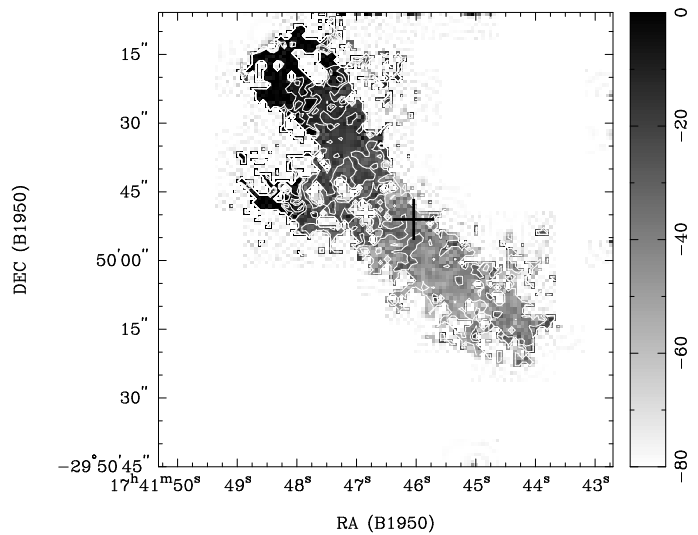}
\caption{The line--centre velocity distribution of the \hone line emission in both greyscale and contours. The greyscale units are in \kms\ and the contours are: $-5, -10, -15, -25, -35$ and $-$55\kms. The cross  marks the location of \oh maser {\em A}.}
\label{fig-vel}
\end{figure}

\begin{figure}
\centering
\includegraphics[height=5cm]{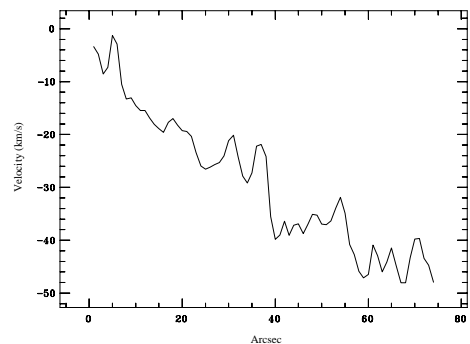}
\caption{A cut, 5 arcsec wide, along the centre of the \h\ bar extending 80 arcsec diagonally from the top of the bar.}
\label{fig-vel_cut}
\end{figure}

The line--centre velocity distribution of H$_2$ 1--0 S(1) emission
(Fig.~\ref{fig-vel}) shows a systematic velocity gradient along the
\h\ bar, with the mean velocity ranging from 0\kms\ in the north to
$-50$\kms\ in the south. These velocities are comparable with the average values of about $-$13\kms\ for the northern part and $-$35\kms\ for the southern part obtained from the repeated observations in June 2000. Fig.~\ref{fig-vel_cut}  shows a cut along the \h\ bar illustrating the velocity gradient.  Note that the small--scale structure showing deviations of $\sim$10--15 \kms\ probably reflects the errors in determining the velocity at each position. 

The presence of a star ($\rm{RA}(1950)=17^{\rm{h}}~41^{\rm{m}}~46 \fs5$ , $\rm{Dec.}(1950)=-29\degr~49\arcmin~45 \farcs0$) in the centre of the H$_2$ filament, together with the velocity difference between the north-east and south-west sections, and the jet--like appearance of the brightest \h\ contours would be consistent with a collimated bipolar jet originating from the star. However, other considerations suggest that this is unlikely. Firstly, the star in question cannot be a protostar. The 2MASS point source catalogue gives the J, H and K magnitudes of the star as $13.22 \pm 0.04$, $10.31 \pm 0.05$ and $8.97 \pm 0.04$ respectively. Using the JHK intrinsic colours from \citet{bb88} and \citet{koornneef83}, and  the reddening vector from \citet{bb88}, we find that the extinction in K-band is $A_K \approx 2 $ (visual extinction $A_V \approx 20 $). These values are consistent with the lack of detection in the Digitized Sky Surveys (DSS). After correcting for extinction, we find that derived magnitudes are consistent with either a K dwarf within 100 pc of the Sun, or a K or M giant at a distance of several kpc.  The former case can be excluded given that there are no molecular clouds within 200~pc of the Sun that could give rise to the H$_2$ emission.  In the latter case the bright star is evolved and cannot be driving the outflow.  The apparent location of the star at the centre of the \h\ bar is, therefore, most probably a chance alignment. Of course, it is possible that apparent outflow is being produced by a star at several kpc which is too faint to be detected.  A more serious problem is that the emission does not resemble that of other outflows detected in the 1--0 S(1) line. In bipolar outflows, such as Cep E \citep{eis96}, the emission associated with jets generally peaks in discrete knots or lobes at some distance from the driving source. Furthermore,  the jets are narrow close to the driving source and the lobes have distinct, roughly constant velocities.  The velocity gradient and morphology, including the extended diffuse component, present in our \h\ data are thus not consistent with entrainment by a fast, narrow protostellar jets. 

\begin{figure}
\includegraphics[height=6.5cm]{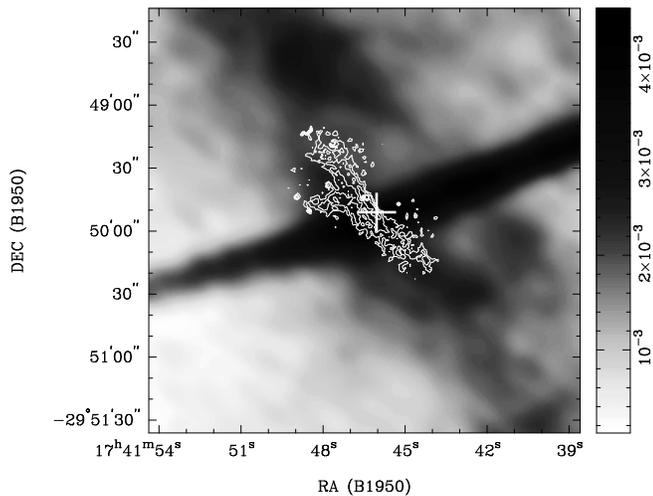}
\caption{The velocty--integrated image of the \h\ line emission  (contours) and the 6--cm radio continuum image (greyscale). The H$_2$ filament is located where the Snake (the dark east--west bar) crosses the shell of G359.1--0.5 (dark band extending almost north-south).  The greyscale units are in Jy\,beam$^{-1}$ and the contours are: 13, 20, 30, 45, 56 and 65 \ee{-6}\ergs. The cross  marks the location of \oh maser {\em A}.}
\label{fig-h2+radio}
\end{figure}

Fig.~\ref{fig-h2+radio} shows contours of velocity--integrated \h\ line emission superimposed on a 6--cm greyscale VLA image with a resolution of 
12 \farcs7 $\times$ 8 \farcs2 (P.A. = 62\degr) \citep{yz01-2}. The bar of \h\ emission is almost perpendicular to the Snake and it runs parallel to the edge of an elliptical depression located $\sim$15 arcsec to the west of the \h\ bar. The synchrotron emissivity of this depression is reduced by $\la 30$ per cent and it surrounds the Snake filament where it crosses the shell of G359.1--0.5.

\subsection{Characteristics of associated molecular cloud} 
 
\subsubsection{Identifying the cloud associated with G359.1--0.5}

\begin{figure}
\rotatebox[origin=cc]{-90}{\includegraphics[height=6cm]{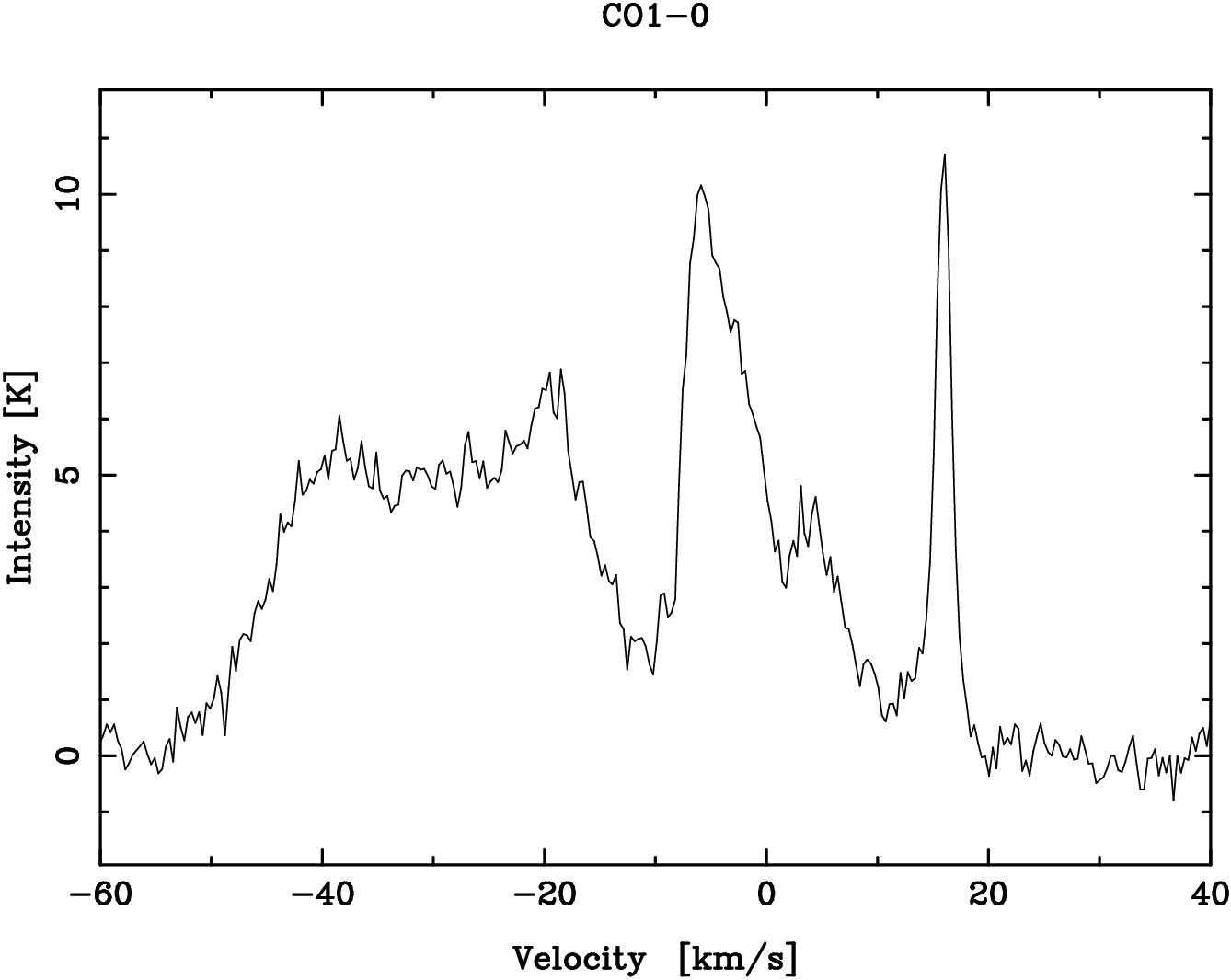}}
\rotatebox[origin=cc]{-90}{\includegraphics[height=6cm]{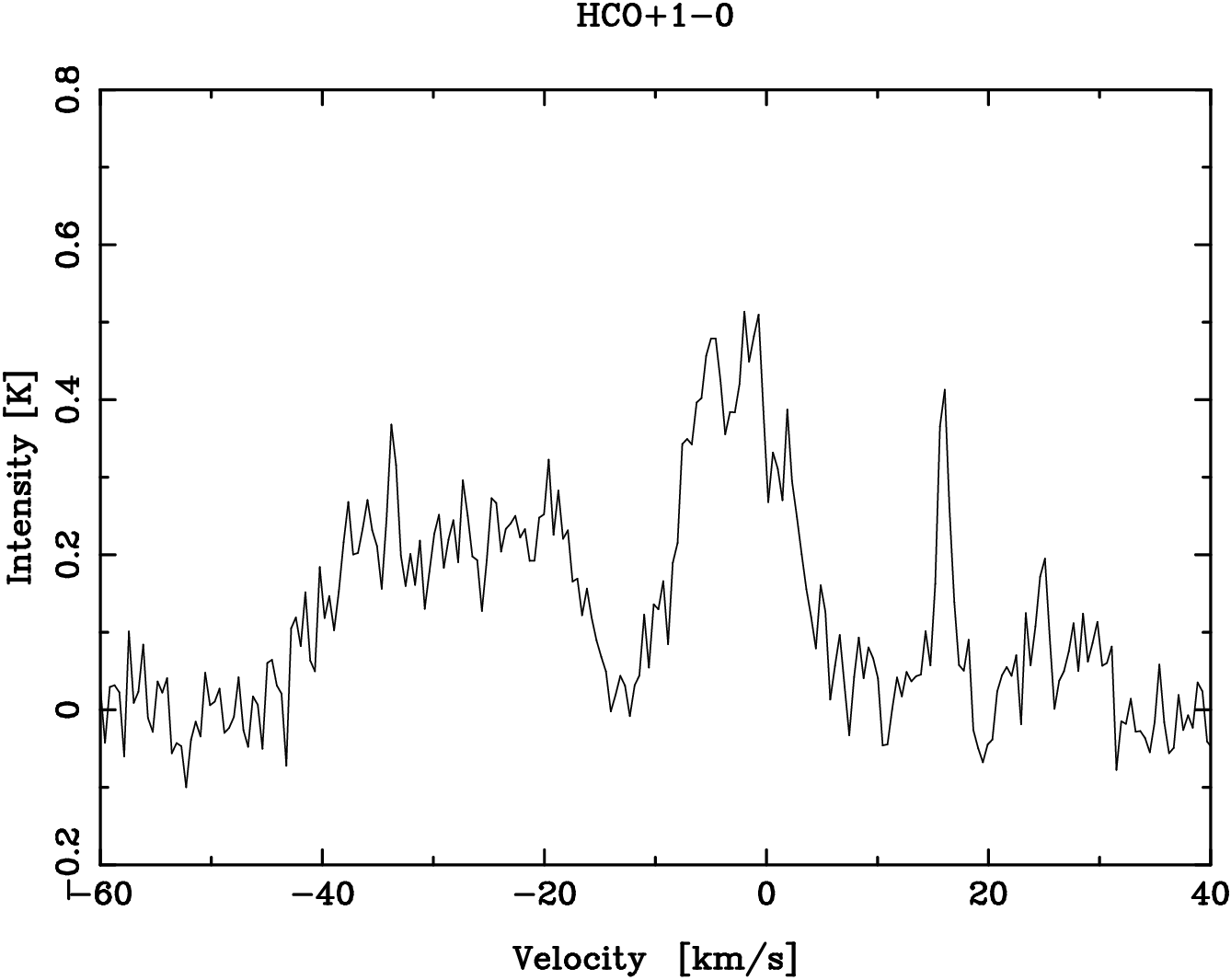}}\\
\rotatebox[origin=cc]{-90}{\includegraphics[height=6cm]{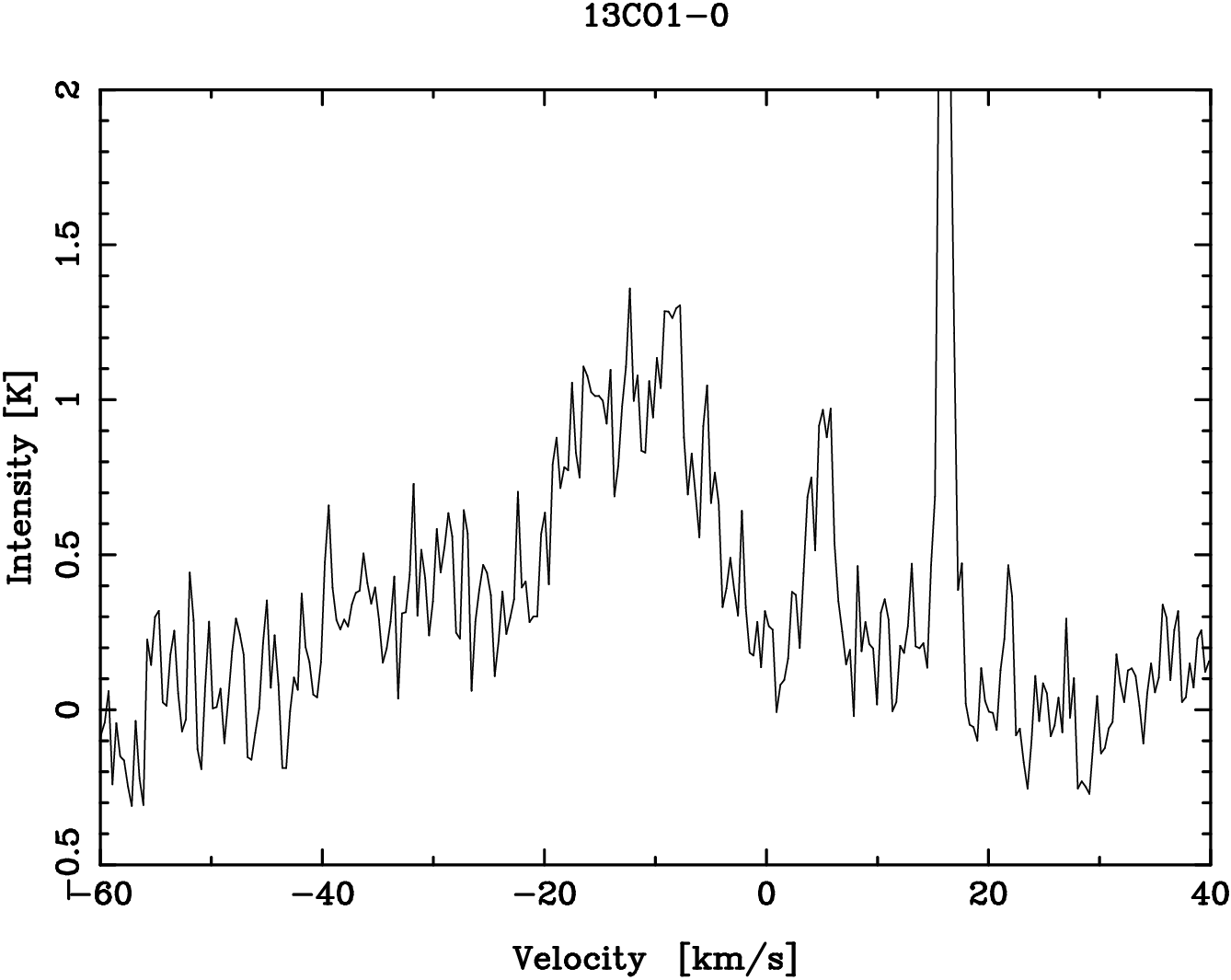}}
\rotatebox[origin=cc]{-90}{\includegraphics[height=6cm]{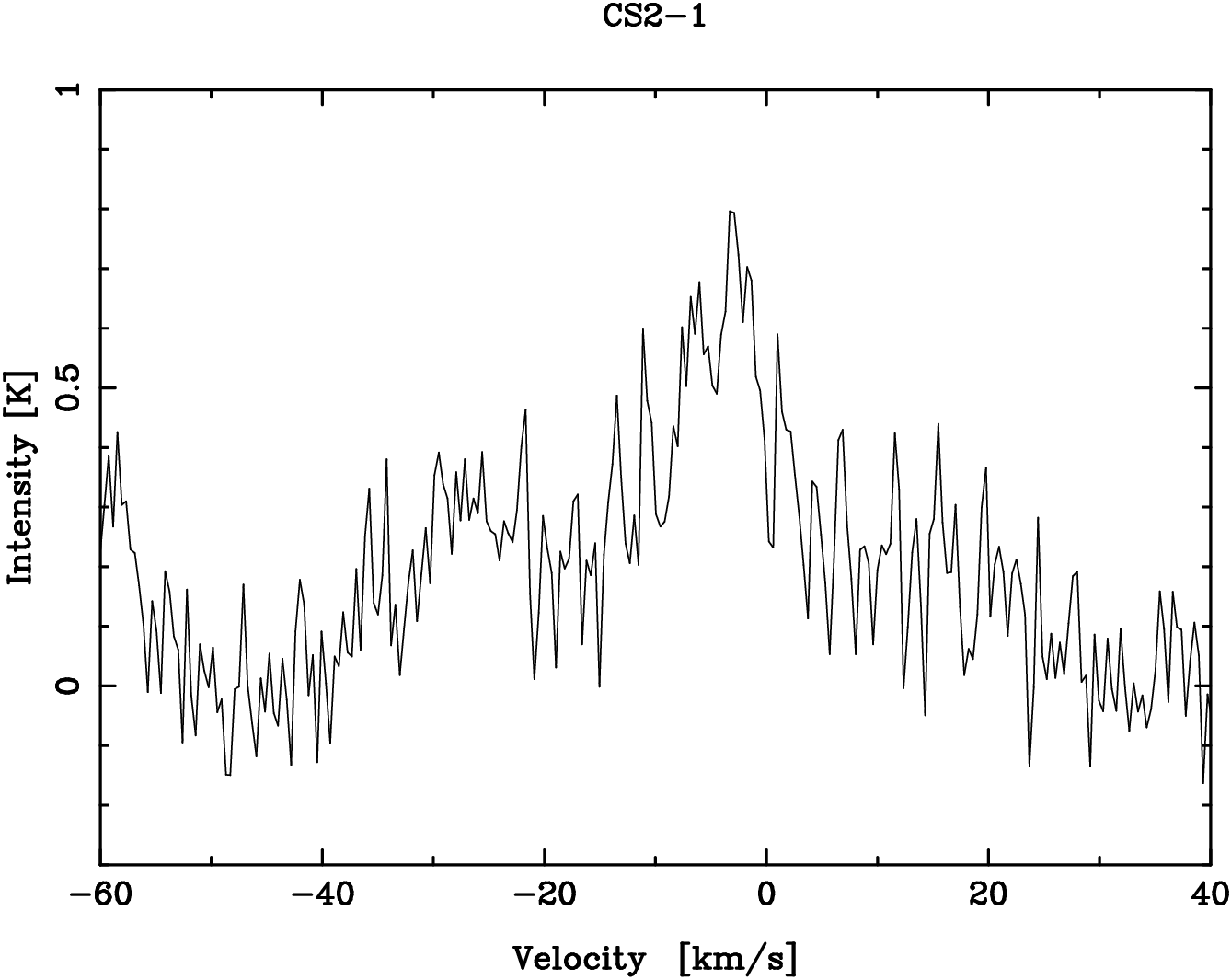}}
\caption{\co, \hcop, \tco\ J=1--0 and CS J=2--1 spectra towards maser {\em A}.}
\label{fig-co}
\end{figure}

Observations of molecular clouds associated with SNRs are potentially  subject to confusion due to other molecular clouds along the line of sight.  However, the maser velocity of $-$5\kms\ can be related to the systemic velocity of the post--shock material next to the SNR shell because the \oh masers are only seen in orientation where they are transverse to the motion of the expanding shock front. Hence they closely match the systemic velocity of the SNR rather than any local intrinsic expansion. We therefore expect that molecular gas with similar velocity originates from the molecular cloud associated with G359.1--0.5. In Fig.~\ref{fig-co} we show the spectra of \co, \hcop, \tco\ J=1--0 and CS J=2--1  towards maser {\em A}. In the velocity range of interest, between $-$20 and 0\kms, the shape of the \tco\ line profile differes from those of the other profiles. Inspection of the velocity cubes implies that emission within this range is most likely associated with the SNR shell.  In contrast, the emission between $-$40 and $-$20\kms\ and the narrow features at $+$5 and $+$16\kms\  appear to be uncorrelated with the SNR shell or the \h\ bar. 

Compared with the \tco\ spectra, the \co~ and \hcop\ spectra, and to a lesser extent the CS spectra, all show an absorption feature between $-$20 and $-5$\kms. In Table~\ref{tab-sest} we list the results of a Gaussian fit to the  \tco\ feature, but for the other species affected by absorption we give only the maximum of the observed line profiles, which thus represent minimum \tmb\ for these species. The correspondence between the peak of the \tco\ emission and the centre of the absorption dip in the other three spectra implies that part of the molecular cloud is being self--absorbed in the molecular species with higher optical depth.

\begin{figure}
\includegraphics[height=5cm]{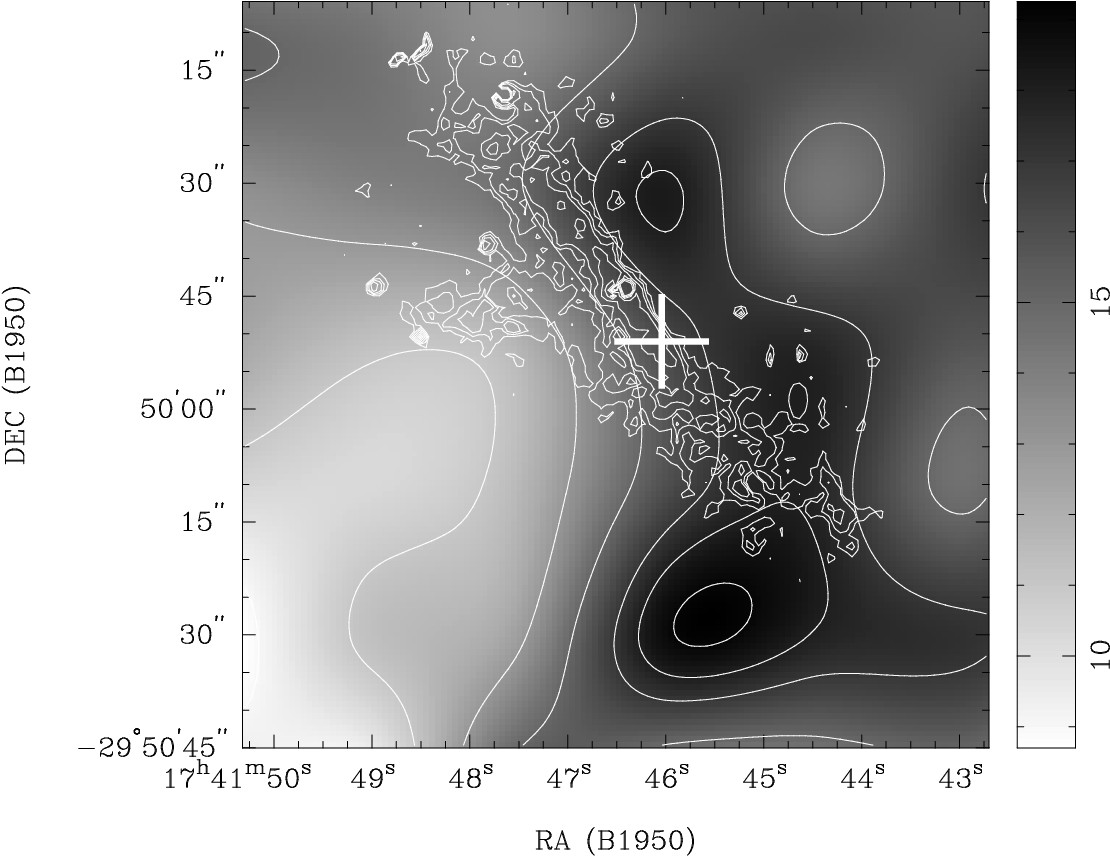}
\includegraphics[height=5cm]{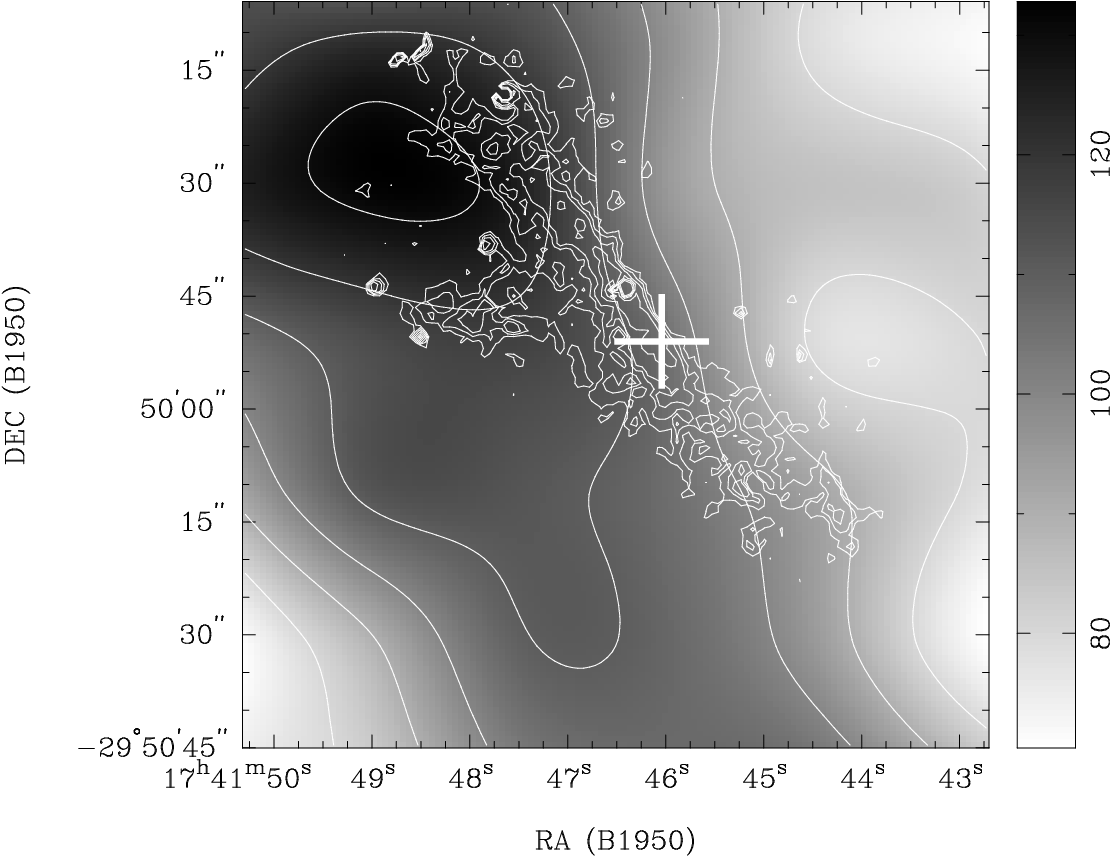}
\includegraphics[height=5cm]{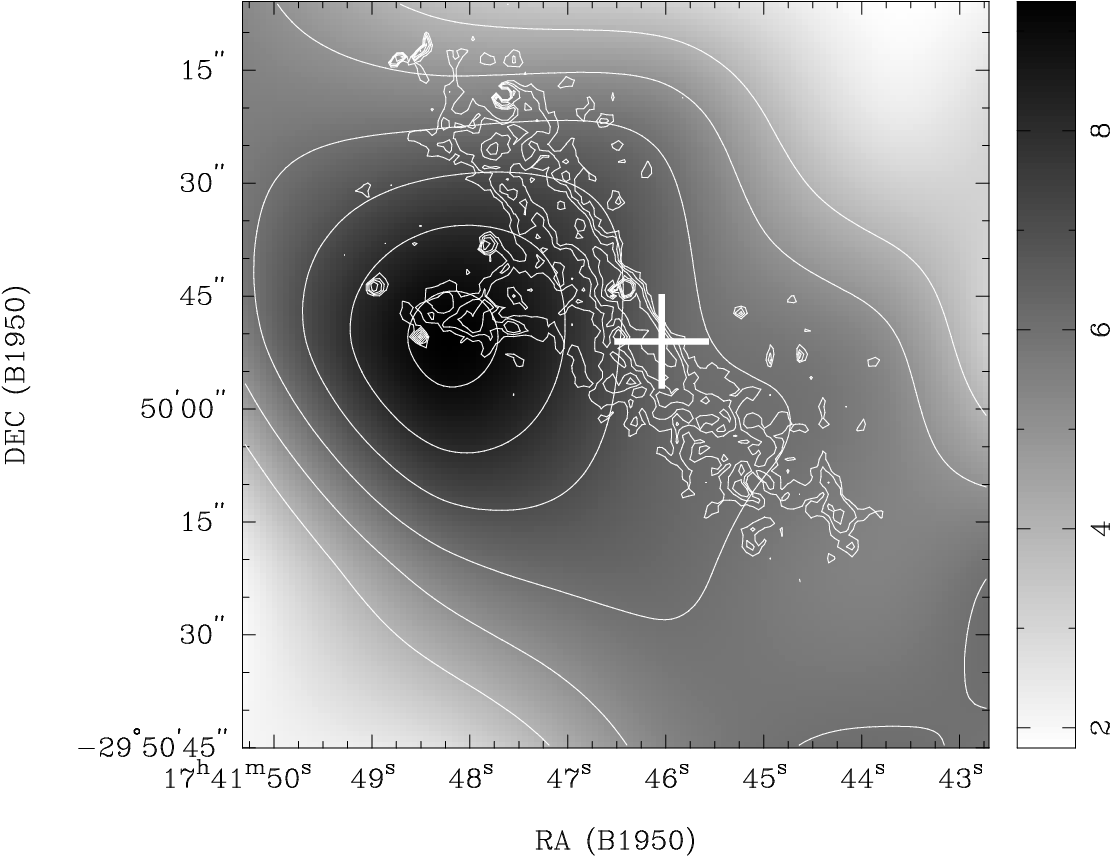} 
\caption{Maps of \tco\ J=1--0 (upper), \co~ J=1--0 (middle) and CS J=2--1 (lower) emission (shown in greyscale with matching contours) integrated between $-20$ and 0\kms\ overlaid with the contours of H$_2$ emission.  The \h\ contours are: 13, 20, 30, 45, 56 and 65 \ee{-6}\ergs. Contour levels (in K\kms) are: (\tco): 9, 11, 13, 15, 17, 18 and 19; (\co): 70, 80, 90, 100, 110, 120 and 130; (CS): 4, 5, 6, 7, 8 and 9. The cross  marks the location of the \oh maser {\em A}.}
\label{fig-int+h2}
\end{figure}

The overall distribution of \co, \tco\ and CS emission show
differences due to variations in optical depth across the region (the
spectra of \hcop\ were not obtained over the whole region, as the
other molecules, but only towards the maser and in four positions around the maser). Maps of the emission, integrated  between $-20$ and 0\kms,  are shown in  Fig.~\ref{fig-int+h2} overlaid with the contours of H$_2$ emission. The true distribution of gas column density is represented by the \tco\ because of its low optical depth, in contrast to the high optical depths of the other species. We note that the \tco\ distribution is dominated by the gas located to the west of the \h\ bar. Although the \co~ and CS emitting gas is covering the same region as shown by the \tco\ map, these molecules show regions dominated by optically thin gas which is not self--absorbed, and is located east from the \h\ bar. East of the bar we thus have the gas with lower optical depth, while to the west the gas has higher optical depth. The \co~ and CS maps both show an elongated feature near and parallel to the \h\ emission and the SNR shell. The \h\ bar is located at the edge of this filament of molecular gas. 

\subsubsection{Physical properties of the molecular cloud}

Self--absorption in molecular clouds can be produced by an overlaying
colder or sub--thermally excited layer. To estimate the  kinetic
temperature and density of the gas we have modeled the observed values
of the \tco\ J=2--1 and J=1--0 line intensities using a statistical--equilibrium excitation code supplied  by J. H. Black. The code employs a  mean--escape probability (MEP) approximation for radiative transfer \citep{jan94} and calculates line intensities given kinetic temperature and density of the gas, the total column density of the molecule and the line width. For the western part of the cloud we find gas temperature of 8 $\pm$ 2~K, gas density of \e{4}\cm{-3}, \tco\ column density of 2.0\ee{16}\cm{-2} and \tco\ J=1--0 optical depth of 0.3. This low optical depth is consistent with the non--detection of C$^{18}$O transitions. Also, the density of \e{4}\cm{-3} is consistent with the molecular distribution being affected by absorption in this part of the cloud, as the lower--excitation temperatures occur at these densities for the \hcop\ and CS lines. Using a \co/\tco\ abundance ratio of 30 \citep{whiteoak82} we have the \co~ column density of 6.0\ee{17}\cm{-2} and \co~ J=1--0 optical depth of 9. On the other hand, the eastern part of the cloud must have density $\ge$ \e{5}\cm{-3} because of the CS J=3--2 detection. For such densities the corresponding gas temperature derived from the code is 10 $\pm$ 2~K, \tco\ column is \e{16}\cm{-2} and \tco\ J=1--0 optical depth is 0.1. This yields  a \co~ column density of 3.0\ee{17}\cm{-2} and optical depth of 3.

These results show that western region of the cloud may be slightly colder than the eastern region. To investigate whether this difference in temperature is sufficient to produce the observed self--absorption, we use the following:
\begin{equation}
T_{abs}= f(T_{bg})e^{- \tau_{fg}}+ f(T_{fg})(1-e^{- \tau_{fg}}),
\end{equation}
where
\begin{equation}
f(T)=\frac{h\nu}{k}[(e^{\frac{h\nu}{kT}}-1)^{-1} - (e^{\frac{h\nu}{kT_{cmb}}}-1)^{-1}].
\end{equation}
$T_{abs}$ is the observed temperature at the velocity of maximum
absorption in \co, $T_{fg}$ is the excitation temperature of the
foreground part of the cloud, $T_{bg}$ is the  excitation temperature
of the background part of the cloud and $\tau_{fg}$ is the optical
depth of the foreground cloud. The observed  $T_{abs}$ ranges from 2
to 3.3~K across the mapped region and $T_{cmb}$ = 2.7~K is the
temperature of the cosmic microwave background. Using $T_{bg}$ = 10~K
and $\tau_{fg}$ = 3 for the eastern part of the cloud we derive the
foreground cloud excitation temperature of $T_{fg} = 6 \pm 1$~K. For
the western part of the cloud  $T_{abs}$ depends only on the
excitation temperature of the foreground part of the cloud because of
the high optical depth in that region ($\tau_{fg}$ = 9). Fitting the observed $T_{abs}$ we derive same $T_{fg}$ of 6 $\pm 1$~K. 

The kinetic temperature of the western part of the cloud (8 $\pm$ 2~K) is in good agreement with the excitation temperature of 6 $\pm$ 1~K derived for the foreground part of the cloud. In other words, for the western part of the cloud \tkin = \tex. Since this is the part of the cloud where \co~ distribution is affected by absorption, we conclude that physically cold, not sub--thermally excited layer of the cloud is responsible for the self--absorption in the observed spectra.

\section{Discussion}

The \h\ 1--0/2--1 S(1) line  ratio of $\sim$ 20  is consistent with shock
excitation, as expected from the presence of the \oh maser. Although
UV excitation can, in principle, produce the 1--0 S(1) line intensity
without violating the limit on the 2--1 line when molecular gas of
density \e{5} -- \e{6}\cm{-3}  is exposed to a FUV flux of $G_0 \ga
10^{4.5}$ \citep{bur90}, there is no OB star within 1~pc, as evident
from the 2MASS point source catalogue, there are no H\,{\sc ii}
regions apparent in the continuum map, and no significant emission in {\it the Midcourse Space Experiment}\footnote{http://www.ipac.caltech.edu/ipac/msx/msx.html} ({\it MSX}) bands at 8.3 and 21.3 \micron\ within an arcmin of the maser.

The \h\ emission must then originate from a shock driven by the
SNR. The \oh maser is expected to be located within the cooling gas
behind a C--type shock wave \citep{lock99}, and its location just
inside the sharp western edge of the \h\ emission supports the idea
that the expanding shell of G359.5-0.1 is driving a shock into a
molecular cloud at that location.  Shock models {\it do not} predict OH
column density required to produce \oh masers
\citep{draine83,kaufman96}. \citet{lock99} and \citet{wardle99}
suggested that the OH column density is enhanced by UV
photodissocaition of H$_2$O. The internal FUV field  generated by the
X--rays from an SNR interior are able to dissociate more than 1 per
cent of H$_2$O, which is enough to produce the required OH abundance
\citep{wardle99}. How does this FUV field affects the \h\ emission? It
is about 100 times weaker than the standard interstellar field
(1.6\ee{-3}\erg), which makes it far less than that associated with an H\,{\sc ii} region or PDR. The corresponding local X--ray energy deposition per
particle is then $H_x/n \sim 7\times10^{-29}$\,erg\cm{3}\,s$^{-1}$
(see equation 2 from \citealt*{maloney96}). From Fig.6a in
\citet{maloney96} the intensity in the \hone line
of 3.2\ee{-5}\ergs\ which is 0.2 per cent of observed peak \h\
flux. Thus, the FUV field produced by X--rays has negligible contribution
to the excitation of the \hone line.

The \h\ bar follows the distribution of extended \oh maser emission \citep{yz95}, which is believed to be an evidence of this interaction on the global scale \citep{yz99}. Indeed, X--ray observations \citep{bamba00} show a centre--filled morphology of this SNR which, in conjunction with the shell morphology in radio band, has been proposed as a characteristic of SNRs interacting with molecular clouds \citep{rho98}. We found the \h\ emission near other masers in this SNR aligned well with the SNR shell \citep{lazendic01}. The magnetic field along the line of sight towards maser {\em A} was estimated to be $\sim 560 \mu$G and the magnetic field was found oriented along the \h\ bar \citep{yz01}. The thickness of the shock is expected to be of order \e{17}~cm, about an arcsec at a distance of 8.5~kpc.  We identify the sharp western edge of the bar as being the leading, limb brightened edge of a curved shock front.  The maser velocity is distinct from the mean velocity of about $-$30\kms\ of the shocked \h\ at this location, but the maser is preferentially produced when the shock front is perpendicular to the line of sight. The \h\ velocity gradient could be the result of a pre--existing gradient, since a gradient of $\sim$5\kms\ is found in the \tco\ gas in same direction.

\begin{figure}
\vspace{5.5cm}
\includegraphics[height=8cm]{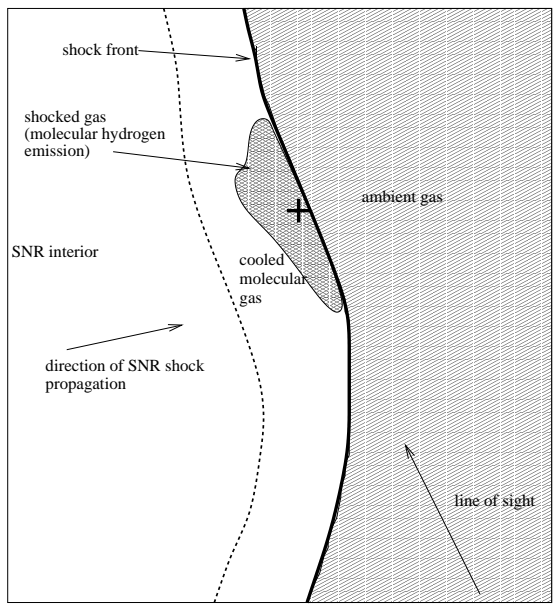}
\caption{A schematic model of the molecular cloud towards the 
maser {\em A} (marked with a cross) is presented to illustrate 
the distribution of different components. The \h\ bar is located 
just behind the thin shock front, which is expanding in a direction 
transverse to the line of sight.}
\label{fig-model}
\end{figure}

The results derived from the molecular spectra imply the presence of a density gradient across the region, and the obvious mechanism to produce such a gradient is the passage of an SNR shock wave. The denser gas ($\sim$\e{5}~-- \e{6}\cm{-3}) then corresponds to post--shocked gas and the lower density gas ($\sim$\e{4}\cm{-3}) corresponds to ambient pre--shocked gas, with the H$_2$  filament being the boundary between the two regions.  A similar case of self--absorption by  pre--shocked gas was also seen in IC 443 \citep{white87, vand93}. The values obtained for gas densities are consistent with those expected in shocks. Because of the presence of the \oh maser,  which requires a density $\sim$\e{5}\cm{-3}, the pre--shock density must be $\sim$\e{4}\cm{-3} since the typical compression in shocks is $\sim$20--30 (for shock speed of 20--30\kms\ and Alfv$\acute{\rm {e}}$n speed in molecular clouds of 2\kms). In Fig.~\ref{fig-model} we present a schematic model of the SNR shock passing through the molecular cloud. Dense gas is located mostly on the eastern side of the \h\ filament, i.e. behind the shock front, while the lower density gas extends mostly to the west from the \h\ emission, i.e. in front of the shock front. 
 
The gas temperature between the pre--shock and post--shock gas does not differ much since the gas cools rapidly behind the C--type shock. We did not detect the warm molecular gas, other than in \h, found in other SNRs associated with \oh masers \citep{vand93,frail98,rr99}. The reason is probably beam dilution. The warm gas is less than 0.6~pc away from the shock front, as inferred from the \h\ emission, and our beam of 1--2 pc in size includes the volume of gas further away that has been cooled to near--ambient gas temperature. The post--shock/ambient cloud temperature of 10~K is consistent with the excitation temperature derived for the clouds in the Galactic Centre region \citep{oka98}. The beam dilution might be also the reason for non--detection of \htco\ and SiO transitions which favour regions with higher temperatures. Although the pre--shock gas temperature of $\sim$8~K is lower than typical molecular cloud temperatures of $\geq 10$~K \citep{gold87}, dark clouds are found with temperatures low as this \citep{dickman75, snell81}. In our case, a likely explanation for this is the broad line width of the cloud (15 -- 20\kms) which enables more efficient cooling. This is supported by the results of the MEP code. As CO is the dominant coolant at densities $\sim$\e{4}\cm{-3} \citep{gold78}, we can estimate the total cooling rate using the total power in the rotational transition of \co. For a \co~ column density of \e{18}\cm{-2} and volume density of \e{4}\cm{-3}, the power emitted for $\Delta v$ = 20\kms\ and  \tkin= 8~K is $\sim$4\ee{-23}\,erg\,s$^{-1}$\,CO$^{-1}$, and for  $\Delta v$ = 2\kms\ and  \tkin= 15~K is $\sim$4.7\ee{-23}\,erg\,s$^{-1}$\,CO$^{-1}$. Thus the cooling rate from a 8~K cloud with 20\kms\ line width is similar to that of a 15~K cloud with a 2\kms\ line width.

Despite the  positional coincidence and spatial proximity of the two, we did not find any morphological evidence in our data that supports the interaction between the SNR and the Snake. Similarly besides the positional coincidence, there is no obvious relationship between the \h\ emission and the Snake. In the view of the recent suggestion by \citet{bi00} that the origin of the Snake is linked to rotating molecular clouds, we examined the possibility that the velocity gradient in H$_2$ of $\pm$20\kms\ might originate from rotation of a flattened cloud. The 1--0 S(1) emission runs along an elliptical depression in the 6--cm emission from the SNR where the Snake crosses it (see Fig~\ref{fig-h2+radio}). This depression could be caused by a rotating dense molecular cloud. However, we derive the mass of the cloud from our \tco\ data to be  $\sim 700$\msol, in which case its escape speed is about 1.6\kms\ and its self--gravity could not sustain the rotation.

\section{Conclusions}

We carried out near--IR and millimetre wavelength observations of the
region around the \oh maser located at the apparent intersection of
the nonthermal filament the Snake and the SNR shell G359.1--0.5. A bar
of \h\ emission was found encompassing the \oh  maser and aligned with the SNR shell. We suggest that \h\ emission originates from the expansion of the SNR blast wave, which is evident from the sharp western edge that corresponds to the forward shock. This shocked \h\ emission, as inferred from the 1--0 and 2--1 S(1) line ratio, supports the notion that the \oh masers association with SNRs are produced in molecular shock waves. 

Emission from the molecular species \co, CS, \hcop\ and \tco\  was detected at the maser velocity. The spectra of the former three species are affected by broad--line absorption which originates from a colder layer of molecular gas. Optically thin  \tco\ was unaffected by the absorption and is produced mostly in this colder layer, identified as pre--shocked gas. The inferred density gradient across the region places higher density post--shocked gas west of the \h\ bar. The distribution of the pre--shock and post--shock gas is consistent with the passage of the SNR shock and the \h\ bar being the boundary between the two regions. 

The warm molecular gas from post--shocked gas was not detected in millimetre wavelengths probably because of beam dilution. For a better morphological and qualitative study of the shocked molecular gas towards maser {\em A} we plan to obtain observations at sub--millimetre wavelengths, which will have improved spatial resolution. Observations at other maser positions in the SNR are in progress to obtain a global view of the interaction of the SNR G359.1--0.5 with its molecular gas environment.

\section*{Acknowledgments}

We thank both the SEST and the AAT for the allocation of observing time.  We thank J. Black for use of his MEP code and M. Cohen, M. Reid and T. Bourke for useful discussions. JSL was supported by the Australian Government International Postgraduate Research Scholarship and the Sydney University Postgraduate Scholarship, and acknowledges travel support from the Access to Major Research Facilities Program of the Australian Nuclear Science \& Technology Organisation. MW was supported by the Australian Research Council. 

This publication makes use of data products from: (1) the Two Micron All Sky Survey (2MASS), which is a joint project of the University of Massachusetts and the Infrared Processing and Analysis Center/California Institute of Technology, funded by the National Aeronautics and Space Administration and the National Science Foundation; and (2) the Digitized Sky Surveys (DSS) were produced at the Space Telescope Science Institute under U.S. Government grant NAG W-2166. The images of these surveys are based on photographic data obtained using the Oschin Schmidt Telescope on Palomar Mountain and the UK Schmidt Telescope. The plates were processed into the present compressed digital form with the permission of these institutions. The National Geographic Society - Palomar Observatory Sky Atlas (POSS-I) was made by the California Institute of Technology with grants from the National Geographic Society.


\end{document}